\documentclass[3p,twocolumn,sort&compress, fleqn]{elsarticle}
\usepackage{amsmath}
\usepackage{amssymb}
\usepackage{bm}
\usepackage{braket}
\usepackage{listings}
\usepackage{cprotect}
\usepackage{lmodern}
\usepackage{ulem}
\usepackage{listings,jvlisting}

\journal{Computer Physics Communications}

\newcommand{\tr}[1]{#1}

\usepackage{natbib}
\usepackage{hyperref}
\hypersetup{
        colorlinks=true,
}

\long\def\beginmypgfpdfnamed#1#2\endmypgfpdfnamed{\includegraphics{#1}}

\graphicspath{{./pyfig/}}

\begin{document}

\begin{frontmatter}

\title{Bayesian optimization package: PHYSBO}

\author[issp]{Yuichi Motoyama}\ead{y-motoyama@issp.u-tokyo.ac.jp}
\author[nims,nims2,gsfs,riken]{Ryo Tamura}\ead{tamura.ryo@nims.go.jp}
\author[issp]{Kazuyoshi Yoshimi}\ead{k-yoshimi@issp.u-tokyo.ac.jp}
\author[riken,yokohama]{Kei Terayama}
\author[gsfs]{Tsuyoshi Ueno}
\author[nims2,gsfs,riken]{Koji Tsuda}

\address[issp]{Institute for Solid State Physics, University of Tokyo, Chiba 277-8581, Japan}
\address[nims]{International Center for Materials Nanoarchitectonics (WPI-MANA), National Institute for Materials Science, 1-1 Namiki, Tsukuba, Ibaraki 305-0044, Japan}
\address[nims2]{Research and Services Division of Materials Data and Integrated System (MaDIS), National Institute for Materials Science, 1-1 Namiki, Tsukuba, Ibaraki 305-0044, Japan}
\address[gsfs]{Graduate School of Frontier Sciences, The University of Tokyo, 5-1-5 Kashiwanoha, Kashiwa, Chiba 277-8561, Japan}
\address[riken]{RIKEN Center for Advanced Intelligence Project, 1-4-1 Nihonbashi, Chuo-ku, Tokyo 103-0027, Japan}
\address[yokohama]{Graduate School of Medical Life Science, Yokohama City University, Yokohama 230-0045, Japan}

\begin{abstract}
PHYSBO (optimization tools for PHYSics based on Bayesian Optimization) is a Python library for fast and scalable Bayesian optimization. 
It has been developed mainly for application in the basic sciences such as physics and materials science.
Bayesian optimization is used to select an appropriate input for experiments/simulations from candidate inputs listed in advance in order to obtain better output values with the help of machine learning prediction.
PHYSBO can be used to find better solutions for both single and multi-objective optimization problems.
At each cycle in the Bayesian optimization, a single proposal or multiple proposals can be obtained for the next experiments/simulations.
These proposals can be obtained interactively for use in experiments.
PHYSBO is available at \url{https://github.com/issp-center-dev/PHYSBO}.
\end{abstract}


\end{frontmatter}
{\bf PROGRAM SUMMARY}

\begin{small}
\noindent
{\em Program Title:} PHYSBO \\
{\em Journal Reference:}                                      \\
{\em Catalogue identifier:}                                   \\
{\em Licensing provisions:} GNU General Public License version 3\\
{\em Programming language:} Python3 \\
{\em Computer:} PC\\ 
{\em Operating system:} Any, tested on Linux and macOS
{\em External routines/libraries:} NumPy, SciPy, MPI for Python. \\
{\em Nature of problem:} 
 {Bayesian optimization (BO) can be used to select inputs that will yield better outputs from a list of candidate inputs with the help of machine learning prediction through a Gaussian process. Although BO is a powerful tool, 
 two of its components, training the Gaussian process regression and optimizing the acquisition function, are generally computationally expensive. Moreover,  hyperparameter tuning is necessary for the former process.
 }
\\
{\em Solution method:} 
 {PHYSBO is a Python library for performing fast and scalable Bayesian optimization. To avoid the computationally expensive training process, PHYSBO uses a random feature map, Thompson sampling, and a one-rank Cholesky update.
In addition, PHYSBO performs hyperparameter tuning automatically by maximizing the Type II likelihood, and MPI parallelization is used to reduce the calculation time for optimizing the acquisition function.}
\\
\end{small}


\section{Introduction}
Optimization problems are among the most important and commonly encountered problems in the basic sciences, such as physics and materials science, in terms of both experiments and simulations. 
In such problems, appropriate inputs need to be chosen in order to produce better outputs.
For example, in material development,
the input is the composition of the elements and the process of fabrication, while the outputs are the material properties.
Bayesian optimization (BO) can be used to select inputs that will yield better outputs from the candidate inputs listed in advance with the help of machine learning prediction through a Gaussian process (GP)~\cite{GPML2005, Candela2005, Terayama-2021}.
In  the field of physics and materials science, successful results from BO have been reported for a broad range of applications, including thermal conductivity~\cite{Shenghong7.021024},
Li-ion conductivity~\cite{Homma9b11654},
epitaxial TiN thin film~\cite{OHKUBO2021100296},
powder manufacturing~\cite{TAMURA2021109290},
scattering experiments~\cite{Noack:2019vy,Vargas_Hern_ndez_2019}, crystal structure~\cite{PhysRevMaterials.2.013803}, fluid dynamics~\cite{Tran2019}, and effective model estimation\cite{Tamura0193785}.

\begin{figure}[t]
  \begin{center}
    \includegraphics[width=0.48\textwidth]{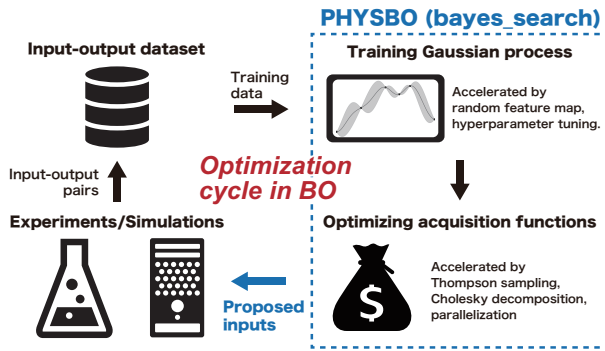}
  \end{center}
  \caption{Flow of Bayesian optimization  {cycle} to find inputs with better output.
  }
  \label{fig:flow_BO}
\end{figure}
In BO, a GP that predicts expected output values and their variances is first trained from already-observed input-output pairs.
Next, the trained GP selects the input with the highest probability of obtaining a better output value based on the acquisition function using the expected output and variance.
The true output value for the selected input is then obtained by experiment/simulation. 
BO repeats these processes to identify better inputs (see Fig.~\ref{fig:flow_BO}).
Although BO is a powerful tool, two of its elements are generally computationally expensive: training the GP regression and optimizing the acquisition function~\cite{Tran2019}.
Furthermore, hyperparameter tuning is necessary for the former process, which can be troublesome for users. 

PHYSBO (optimization tools for PHYSics based on Bayesian Optimization) is a Python library for fast and scalable Bayesian optimization.
PHYSBO is a fork project of another BO package, COMBO (COMmon Bayesian Optimization) library~\cite{UENO201618}.
In COMBO, to avoid the high computational expense,
a random feature map~\cite{Rahimi-2007}, Thompson sampling~\cite{NIPS2011_e53a0a29}, and one-rank Cholesky update~\cite{Gill-1974} are used.
The hyperparameter tuning
is performed automatically by maximizing the Type II likelihood\cite{GPML2005}.
COMBO targets discretized inputs. This is because many problems in materials exploration are treated as discretized inputs, and by evaluating the acquisition function of all discretized candidates, the optimization of the acquisition function can be performed rigorously.
Additionally, it makes it easier to impose constraints in inputs.

Continuing this trend, PHYSBO also targets discrete data.
In addition to these features, PHYSBO has following functions:
By using parallelization in optimizing the acquisition function, the time for selecting the next input can be shortened, thus reducing the computational expense of the latter process.
PHYSBO is capable of treating both single- and multi-objective optimization problems.
In the single objective case, a single target output is set and the input with the best target output is searched for.
In the multi-objective case, the optimization deals with multiple targets simultaneously~\cite{Tamaki-1996,Tamaki-2006,Coello-2006}.
In these multi-objective problems, multiple Pareto solutions are searched for. 
Each Pareto solution is an optimal solution obtained by varying the balance of the types of target outputs.
In each cycle of BO, a single proposal or multiple proposals are obtained for the next experiments/simulations.
For some situations, BO must be performed sequentially based on the output obtained from the experiments for the selected inputs. Importantly, PHYSBO is designed to run interactively so that it can be executed in such situations.
The mean value and the variances of the posterior distribution of obtained GP at arbitrary points can be evaluated in PHYSBO.

The remainder of the paper is organized as follows:
In Sec. 2, we briefly introduce the BO algorithm.
In Sec. 3, the basic use of PHYSBO is explained.
Section 3.3 is devoted to describing new functionalities of PHYSBO.
In Sec. 4, results of a PHYSBO application are shown.
Following a discussion of the scalability of BO in Sec. 4.1, examples of materials screening and effective model estimation are shown in Secs. 4.2 and 4.3.
Section 5 provides a brief summary.


\section{Algorithm}\label{sec:algorithm}

In PHYSBO, BO is used to effectively select the next input for the experiments/simulations from the set of candidate inputs prepared in advance.
The candidate inputs are denoted as $\{ \mathbf{x}_i \}_{i=1,...,N}$,
where $\mathbf{x}_i$ is a $d$-dimensional vector and $N$ is the number of candidates.
The value of the target output is written as $y_i$ when $\mathbf{x}_i$ is input.
As initial training data,
$M$ experiments/simulations are performed,
and training data $D=\{ \mathbf{x}_k, y_k \}_{k=1,...,M}$ are prepared.
BO selects the next $M+1$-th data value, $\mathbf{x}_{M+1}$ with the help of machine learning.
After evaluating the value of the objective function, $y_{M+1}$, the $M+1$-th pair $(\mathbf{x}_{M+1}, y_{M+1})$ is added to the training dataset.
Then, this cycle continues (see Fig.~\ref{fig:flow_BO}).

\subsection{Acceleration of calculation in Bayesian optimization}
\label{subsec:Acceleration}
In PHYSBO, a random feature map, Cholesky decomposition, and Thompson sampling are used to accelerate the calculations required for the Bayesian optimization. The details of these methods are described below.

\subsubsection*{Random feature map}
In GP, when training data $D$ are given, $y$ is generated according to the following Gaussian distribution:
\begin{align}
p(y|D) = {\mathcal N} (\bm{\mu}, \Sigma),
\end{align}
where $\boldsymbol{\mu}$ is the expected value of $\mathbf{y} = (y_1, y_2, \cdots, y_M)^\top$, i.e. $\mathbb{E}\left[\mathbf{y}\right]$, and $\Sigma$ is the covariance matrix, i.e. $\mathbb{E}\left[\mathbf{y}\mathbf{y}^\top\right]$.
When the Bayesian linear regression model is used, $y$ is given as 
\begin{eqnarray}
y (\mathbf{x}) = \mathbf{w}^\top \boldsymbol{\phi}(\mathbf{x}) + \epsilon,
\label{eq:linear_model}
\end{eqnarray}
where $\mathbf{w}$ is the coefficient vector, $\boldsymbol{\phi} (\mathbf{x})$ is the feature vector of ${\bm x}$, and $\epsilon$ is noise generated according to $\mathcal{N} (0,\sigma)$.
Then, $\boldsymbol{\mu}$ and $\Sigma$ are given as
\begin{align}
\boldsymbol{\mu} &= (\Phi \Phi^\top + \sigma^2 I)^{-1} \Phi \mathbf{y}, \label{eq:mu}\\
\Sigma &= \sigma^2 (\Phi \Phi^\top + \sigma^2 I)^{-1},\label{eq:sigma}
\end{align}
where $\Phi=(\boldsymbol{\phi}({\mathbf{x}}_1), \boldsymbol{\phi}({\mathbf{x}}_2), \cdots, \boldsymbol{\phi}({\mathbf{x}}_M))$ with the number of known data values $M$. 

The random feature method allows us to approximate the kernel function with a positive definite symmetric function by probabilistic sampling\cite{NIPS2007_013a006f}.
Using this method, 
we can approximate the Gaussian kernel $k(\mathbf{x},\mathbf{x}')$ as
\begin{align}
k(\mathbf{x},\mathbf{x}') &= \exp \left[ - \frac{1}{2 \eta^2} \| \mathbf{x} -\mathbf{x}' \|^2 \right] \nonumber\\
&\simeq \boldsymbol{\phi} (\mathbf{x})^\top \boldsymbol{\phi}(\mathbf{x}'),
\end{align}
where $\boldsymbol{\phi} (\mathbf{x})$ is an $\ell$-dimensional feature vector defined as 
\begin{align}
\boldsymbol{\phi} (\mathbf{x}) = \left( z_{\boldsymbol{\omega}_1, b_1} (\mathbf{x}/\eta),..., z_{\boldsymbol{\omega}_\ell, b_\ell} (\mathbf{x}/\eta) \right)^\top.
\end{align}
Here, $ z_{\boldsymbol{\omega}, b} (\mathbf{x}) = \sqrt{2} \cos (\boldsymbol{\omega}^\top \mathbf{x}+b)$ and a $d$-dimensional vector $\boldsymbol{\omega}$ is generated according to the probability $p(\boldsymbol{\omega}) = (2\pi)^{-d/2} \exp (-\|\boldsymbol{\omega}\|^2/2)$,
and $b$ is selected randomly in $[0, 2 \pi)$.
This approximation is strictly valid in the limit of $\ell \to \infty$.
While random feature maps for some of the other kernels such as the Laplace kernel are known~\cite{NIPS2007_013a006f}, only that of the Gaussian kernel is implemented in PHYSBO.

Note that in the GP, there are two hyperparameters, $\eta$ and $\sigma$; PHYSBO automatically determines these values by maximizing the Type II likelihood~\cite{GPML2005}.

\subsubsection*{Cholesky decomposition}
As seen from Eqs.(\ref{eq:mu}) and (\ref{eq:sigma}),
calculation of the inverse matrix is needed to obtain the posterior distribution function. By using the matrix $A =\Phi \Phi^\top/\sigma^2 +I$, the posterior distributions are expressed as
\begin{align}
p(y|D) &= p(\mathbf{w}|D)\nonumber\\ 
&= \mathcal{N} \left( \boldsymbol{\mu} = \frac{1}{\sigma^2} A^{-1} \Phi \mathbf{y}, \Sigma = A^{-1} \right). \label{eq:post}
\end{align}
Here, we consider the case where $(\mathbf{x}', y')$ is newly added to the training data in an iteration of BO.
With the addition of this data, the matrix $A$ is updated as
\begin{eqnarray}
A' = A + \frac{1}{\sigma^2} \phi (\mathbf{x}') \phi (\mathbf{x}')^\top.
\end{eqnarray}
This update can be performed using the Cholesky decomposition ($A= L^\top L$), with a reduced $A^{-1}$ computing cost of $O(\ell^2)$.
If we were to compute $A^{-1}$ at every step without this update scheme, the computing cost would be $O(\ell^3)$. 

\subsubsection*{Thompson sampling}

Next, by using Thompson sampling~\cite{TS},
the computation time for predicting the output value is reduced to $O(\ell)$.
A coefficient vector $\mathbf{w}^*$ is sampled according to the posterior probability distribution, Eq.~(\ref{eq:post}).
The acquisition function to select the next input is given by
\begin{eqnarray}
\text{TS} (\mathbf{x}) = {\mathbf{w}^*}^\top \boldsymbol{\phi} (\mathbf{x}).
\end{eqnarray}

PHYSBO calculates
the value $\text{TS}(\mathbf{x})$ for all the candidate points with unknown value of objective function and then chooses the $\mathbf{x}$ with the maximum value as the next input.
Since $\boldsymbol{\phi}(\mathbf{x})$ is an $\ell$-dimensional vector, 
the acquisition function can be calculated with $O(\ell)$.

\subsection{Other acquisition functions for single objective optimization}
\label{subsec:Acquisition_single}

In addition to Thompson sampling, two acquisition functions, the probability of improvement (PI)~\cite{PI} and the expected improvement (EI)~\cite{EI}, can be used in PHYSBO to select the next input.
PI represents the probability of exceeding the current maximum value of already calculated $y$, $y_{\text{max}} = \max_k y_k$; that is,
\begin{equation}
\text{PI} (\mathbf{x}) = \mathbb{P}\left[y(\mathbf{x}) > y_\text{max}\right]  = F\left(\frac{\mu_c(\mathbf{x}) - y_\text{max}}{\sigma_c(\mathbf{x})}\right),
\end{equation}
where $F(\cdot)$ is the cumulative distribution function of $\mathcal{N}(0,1)$,
and $\mu_c(\mathbf{x})$ and $\sigma_c(\mathbf{x})$ are the mean and standard deviation of the posterior distribution given by Eq.~(\ref{eq:post}) at $\mathbf{x}$, respectively.
EI is the expected value of how much $y_\text{max}$ updates when $x$ is observed; that is,
\begin{equation}
\begin{aligned}
\text{EI} (\mathbf{x})
&= \mathbb{E}\left[\max(y(\mathbf{x}) - y_\text{max}, 0)\right] \\
&= \sigma_c(\mathbf{x})\left[t (\mathbf{x}) F(t(\mathbf{x}))+f(t(\mathbf{x}))\right],
\end{aligned}
\end{equation}
where $t(\mathbf{x}) = (\mu_c(\mathbf{x})-y_\text{max})/\sigma_c(\mathbf{x})$ and $f(\cdot)$ is the probability density of 
$\mathcal{N}(0,1)$.

If a random feature map is used in BO, the GP training time can be reduced even when EI and PI are used. 
On the other hand, when EI and PI are used, the computing time to select the next input is $O(\ell^2)$, since in this case, the standard deviation must be evaluated.
The advantage of TS is that the computational cost is lower for selecting next candidates, $O(\ell)$.

\subsection{Thompson sampling and acquisition functions for multi-objective optimization}
\label{subsec:multi_thompson}

In multi-objective optimizations,
multiple objectives (i.e., multiple output values) are dealt with simultaneously.
The search is for Pareto solutions, at which no improvement in any objective value can be found without degrading the others.
If there is a trade-off between objectives, multiple Pareto solutions exist.
In PHYSBO, for multi-objective optimization, Thompson sampling and two acquisition functions are available.

In common, the surrogate model by the Gaussian process is trained for each target.
In Thompson sampling,
for each surrogate model,
the sampling is performed from the posterior distribution given by Eq.~(\ref{eq:post})~\cite{Yahyaa2015ThompsonSF}.
The predicted values of the objective function are then evaluated by the surrogate models.
Using these values, the inputs that are expected to be the Pareto solutions are searched for.
Basically, multiple inputs are obtained; one input is then randomly selected for the next experiments/simulations.

In PHYSBO, HVPI (hypervolume-based probability of improvement) and EHVI (expected hyper-volume improvement) are implemented as acquisition functions for multi-objective optimization.
The former is based on the improvement probability of a non-dominated region in a multi-dimensional objective function space. The latter is based on the expected improvement of the non-dominated region.
A detailed definition is provided in Ref.~\cite{Couckuyt:2014up}.
In the PHYSBO package, a maximum of three objective variables is recommended due to the computing time of acquisition functions.
Note that, even if three objective cases, EHVI takes a lot of time compared with TS and HVPI.

\subsection{Multiple proposals by Bayesian optimization}
\label{subsec:multi_proposals}

In materials exploration, it is a common case that evaluation of the objective function is costly, but at the same time, it is often possible to perform high-throughput batch experiments or simulations in parallel for several inputs.
Therefore,
in the optimization cycle, multiple proposals are sometimes desired.
In PHYSBO, multiple inputs can be suggested based on a strategy for roll-out~\cite{Snoek-2012}.
The implemented algorithms are as follows:
(1) Construct a surrogate model from the present known training data $D$ by GP.
(2) Select one input based on the acquisition functions.
(3) Add the selected input to the training data by considering its output value as the value predicted by the surrogate model.
(4) Return to (1) and retrain the surrogate model to select the next input.
Repeating this process produces multiple proposals.

\section{Usage}
\subsection{Install}\label{subsec:install}
PHYSBO is written in Python3 (3.6 or higher required).
Installing PHYSBO requires Cython, which is installed by
\begin{verbatim}
$ python3 -m pip install Cython
\end{verbatim}
Because PHYSBO is registered in PyPI, a public repository of Python software,
users can download and install PHYSBO simply by the following command:
\begin{verbatim}
$ python3 -m pip install physbo
\end{verbatim}
This will automatically install the dependencies NumPy and Scipy.
When users wish to employ MPI parallelization, the mpi4py package is required.

\subsection{Basic usage}\label{subsec:usage}
A small example program (Program~\ref{code:sample}) for maximizing $f(\bm{x}) = -\|\bm{x}\|_2^2$ subject to $\bm{x} \in [-2,2]^2$ can be used to illustrate how PHYSBO is implemented.

\begin{lstlisting}[language=python, label=code:sample, caption=Small example of PHYSBO]
from itertools import product
import numpy as np
import physbo

xs = np.linspace(-2.0, 2.0, num=21)
X = np.array([x for x in product(xs, repeat=2)])

def simulator(actions: np.ndarray) -> np.ndarray:
    return -np.sum(X[actions, :]**2, axis=1)

policy = physbo.search.discrete.policy(X)

policy.random_search(max_num_probes=10,
                     simulator=simulator)

policy.bayes_search(
    max_num_probes=20,
    num_search_each_probe=1,
    num_rand_basis=500,
    interval=5,
    simulator=simulator,
    score="TS"
)

best_fxs, best_actions = \
    policy.history.export_sequence_best_fx()
best_fx = best_fxs[-1]
best_X = X[best_actions[-1], :]
print(f"best_fx: {best_fx} at {best_X}")
\end{lstlisting}

\subsubsection{Preparing candidates}
In applying PHYSBO, a set of all candidate inputs must be prepared in advance in the form of a $N \times d$ matrix, where $N$ is the number of candidates and $d$ is the dimension of the parameters of input vector $\mathbf{x}$.
Here, lines 5 and 6 create a $401\times2$ matrix \texttt{X}.
The index of candidates (the first axis of \texttt{X}) is referred to as ``action''.

\subsubsection{Defining objective function}
``Simulator'' is a function\footnote{Strictly speaking, a simulator is a callable object, and hence an instance of a class with the \texttt{\_\_call\_\_} method can also be a simulator.}  changing an array of actions to an array of the corresponding output values of the objective function, $f([i,j,\dots]) = [y_i, y_j, \dots]$.
Lines 8 and 9 define a simulator representing $f(\bm{x}) = -\|\bm{x}\|^2_2$.
It should be noted that PHYSBO solves a \textit{maximization} problem; if the user wishes to solve a minimization problem, they should change the sign of the objective function.

\subsubsection{Optimizing}
\texttt{physbo.search.discrete.policy} is a class of single objective BO.
The constructor takes an array as the set of candidates (line 11).
The \texttt{random\_search} method chooses \texttt{max\_num\_probes} actions randomly and evaluates the \texttt{simulator}.
The chosen actions and the corresponding output values are automatically stored in \texttt{policy} as the training dataset.

The \texttt{bayes\_search} method (line 16--23) performs the BO for \texttt{max\_num\_probes} steps.
\texttt{num\_search\_each\_probe} indicates the number of actions proposed in one probe (step) (see sec.~\ref{subsec:multi_proposals} and \ref{sec:multi-proposals}).
\texttt{num\_rand\_basis} is the basis number (dimension) of the random feature map, $\ell$.
If \texttt{num\_rand\_basis} is zero, the random feature map will not be used.
\texttt{interval} specifies how often the hyperparameters will be tuned.
If \texttt{interval} is positive, the hyperparameters are optimized both at the beginning and after each \texttt{interval} step; if it is zero, the hyperparameters are modified only at the beginning; otherwise, the hyperparameter-tuning process does not run.
\texttt{score} indicates the type of acquisition function (\texttt{"TS", "PI", "EI"}).
As in the \texttt{random\_search}, \texttt{policy} stores the pairs of proposed actions and corresponding output values and adds them to the training dataset.

\subsubsection{Showing results}
\texttt{policy.history} stores the history of the search process.
The \texttt{export\_sequence\_best\_fx} method of \texttt{policy.history} returns two lists with the length of the number of probes: the first is the history of the best target values; the second is the history of the best action.
Thus, the final elements represent the best result of the entire calculation (lines 25--29).

\subsection{Advanced usage}
In the following subsections, we briefly describe advanced functionalities, which are newly added in PHYSBO.
\subsubsection{Interactive optimization}
Evaluation of the output can sometimes take a very long time when the next input is proposed.
For example, to study the composition dependence of some property of chemical compounds, researchers would need to synthesize substances and measure the property.
For such a case, PHYSBO has an ``interactive'' mode.
Invoked without the \texttt{simulator} argument,
the \texttt{random\_search} and \texttt{bayes\_search} methods simply return an action as a proposal without evaluating the objective function.
By using the proposal, users can evaluate the output externally, outside PHYSBO.
Once the value of the objective function is determined, the action and output value pair can be added to the \texttt{policy} instance via the \texttt{write} method.

\subsubsection{Multi-objective}
As noted, PHYSBO can deal with multi-objective optimization problems.
Here, users can use the \texttt{physbo.search.discrete\_multi.policy} class rather than \texttt{discrete.policy}.
Use of the former is similar to that of the latter.
The main difference is that a simulator must return $N_\text{actions} \times N_\text{objective}$ matrix rather than  $N_\text{actions}$ vector, where $N_\text{actions}$ and $N_\text{objective}$ are the number of proposals at each step and the number of objective functions, respectively.
\texttt{"TS"}, \texttt{"HVPI"}, and \texttt{"EHVI"} are available for \texttt{score}.

\subsubsection{MPI parallelization}
For the proposal of the next action, PHYSBO evaluates the value of the acquisition function for all the remaining candidate points and then finds the point with the maximum value.
In order to speed up this procedure, PHYSBO adopts MPI parallelization via mpi4py~\cite{mpi4py2005,mpi4py2008,mpi4py2011,mpi4py2021} (for details, see section~\ref{Sec:scalability}).
Users can pass an MPI communicator to the constructor of the \texttt{policy} class to activate MPI parallelization as
\begin{verbatim}
policy(X, comm=MPI.COMM_WORLD)
\end{verbatim}

\subsubsection{Use as Gaussian process}
After trained of the Guassian process,
users can evaluate the mean value and the variance of the posterior distribution at arbitrary point, $\mu_c(\mathbf{x})$ and $\sigma^2_c(\mathbf{x})$ as follows:
\begin{verbatim}
policy.get_post_fmean(X)
policy.get_post_fcov(X)
\end{verbatim}
By using $\mu_c(\mathbf{x})$ and $\sigma^2_c(\mathbf{x})$,
users can define other acquisition function such as the upper confidence bound (UCB)~\cite{GP-UCB},
\begin{equation}
\text{UCB}(\mathbf{x}) = \mu_c(\mathbf{x}) + \kappa \sigma_c(\mathbf{x}),
\end{equation}
where $\kappa > 0$ is another hyperparameter.


\section{Applications}\label{Sec:Examples}

\subsection{Investigations of scalability}\label{Sec:scalability}
\begin{figure}[t]
  \begin{center}
    \includegraphics[width=0.48\textwidth]{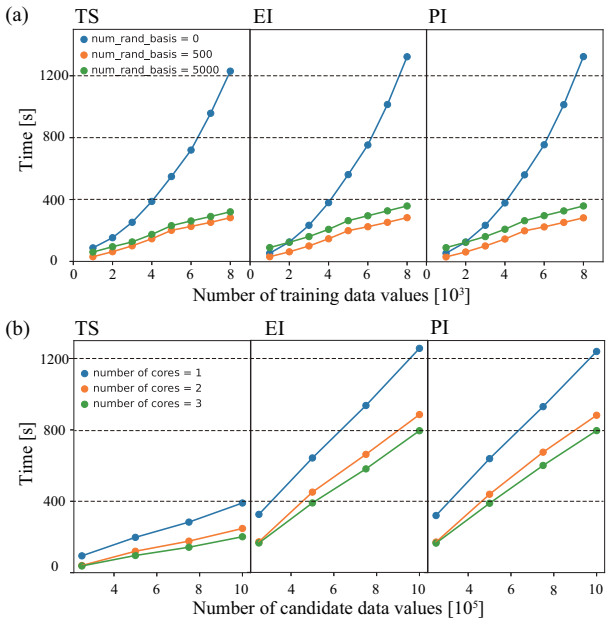}
  \end{center}
  \caption{
(a) Training time as a function of the number of initial data values, which relates to the number of training data values for different scores.
Results when the random feature map is used (\texttt{num\_rand\_basis = 500} and \texttt{5000}) are compared to the results when the random feature map is not used (\texttt{num\_rand\_basis = 0}).
The time needed for hyperparameter tuning to be applied twice is included in the training time. 
(b) Selection time as a function of the number of candidate data values for different scores.
Results using the number of cores are compared.
The time needed for hyperparameter tuning to be applied twice is included in the selection time. 
Computing time is measured using CPU: 2.3GHz Quad Core i7.
  }
  \label{fig:scale}
\end{figure}

The scalability of PHYSBO with respect to the computation time for training and for selecting the next inputs (hereafter referred to as training time and selection time) can be readily demonstrated.
Here, the input is a one-dimensional parameter; the output is a randomly generated real value between 0 and 1.

We first consider the training time with respect to the size of the training data set.
We set the initial number of data values to $N_{\rm i}$ and the number of candidate input values to $N_{\rm i} + 1000$.
We then measure the time to select 200 of the 1000 remaining candidates by PHYSBO (i.e. \texttt{max\_num\_probes = 200}).
The frequency of hyperparameter tuning is set to \texttt{interval = 100},
and single proposal is adopted.
Figure~\ref{fig:scale} (a) plots training time against the initial data values $N_{\rm i}$ for \texttt{score = "TS"}, \texttt{"EI"}, and \texttt{"PI"}.
Also shown is the training time according to the basis number (dimension) of the random feature map, $\ell = 500$ and $5000$, (\texttt{num\_rand\_basis = 500}, \texttt{5000}) and the time taken without a random feature map (\texttt{num\_rand\_basis = 0}).
Irrespective of which score is used, training time is clearly reduced by using the random feature map\footnote{The training time for TS is smaller than PI/EI, but this is because these include the selection time, which of TS is smaller than which of PI/EI.}.
It is also apparent that when a random feature map is used, the training time increases almost linearly with the number of training data values.
It should be noted that because the random feature map approximates the kernel matrix by using random sampling, it works regardless of the number of features $\ell$ and the size of the training data $N$.
However, when $\ell > N$, the overfitting may occur.

The efficiency of parallelization is also of interest.
In order to propose the next input, we need to evaluate the acquisition function for all the not-yet visited candidates.
When one wants to treat the huge candidate space, i.e., wants to enlarge parameter space or to increase the resolution,
more time is obviously needed to evaluate all the values of the acquisition function.
In response, parallelization provides an effective way to achieve faster BO.
PHYSBO adopts the MPI for parallelization using mpi4py~\cite{mpi4py2005,mpi4py2008,mpi4py2011,mpi4py2021}, and each MPI process evaluates the acquisition function for the corresponding part of the entire space.
Here, the initial number of data values is set to $N_{\rm i} = $1,000, and the time to select 200 values is measured in cases where the number of candidates is set to 250,000, 500,000, 750,000, and 1,000,000, respectively.
The basis number in the random feature map is set to \texttt{num\_rand\_basis = 500}, and the frequency of hyperparameter tuning is set to \texttt{interval = 100}.
Figure~\ref{fig:scale} (b) plots selection time as a function of the number of candidates using different cores.
As the number of cores is increased, the selection time decreases, indicating that parallelization is effective for acquisition function optimization.
Furthermore, selection time increases almost linearly with the number of candidates.
Note that the selection time using TS is much lower than that using EI and PI.
Thus, by using TS and multiple cores in a PC, the number of candidates can be increased and large-scale problems become manageable.

\subsection{Materials screening}
The example below shows an application of the BO in PHYSBO to find a material with the desired physical properties of semiconductors.

\begin{figure}[t]
  \begin{center}
    \includegraphics[width=0.4\textwidth]{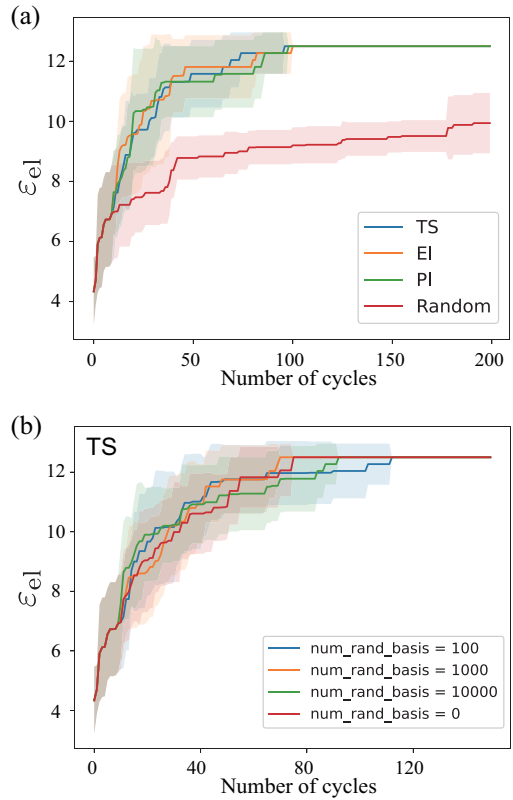}
  \end{center}
  \caption{
(a) Best electronic dielectric constant values ($\varepsilon_\text{el}$) during the optimization as a function of the number of cycles by PHYSBO with \texttt{score = "TS"}, \texttt{"EI"}, and \texttt{"PI"}.
The results are compared with random sampling (Random).
Ten independent runs are performed; the mean and standard deviation are plotted as lines and shaded areas, respectively.
(b) Optimization results as a function of the basis number of the random feature map.
The results are compared with those produced without using a random feature map (\texttt{num\_rand\_basis = 0}).
TS is used to score.
  }
  \label{fig:bandgap_single_probe}
\end{figure}

\begin{figure}[t]
  \begin{center}
    \includegraphics[width=0.4\textwidth]{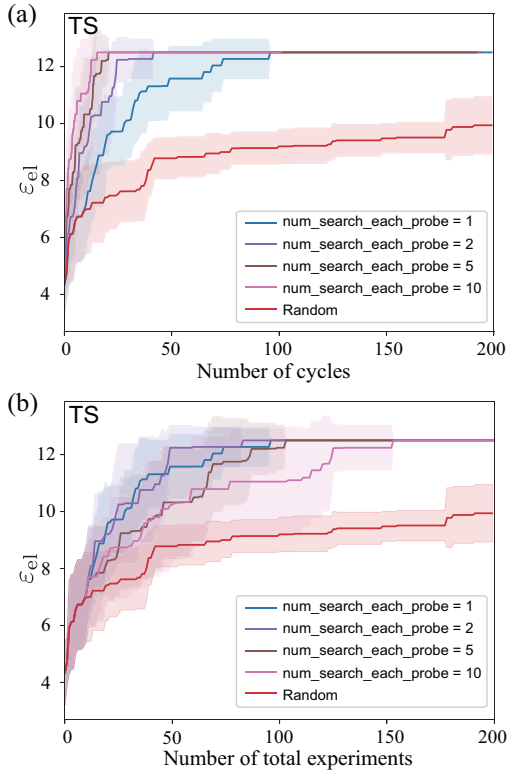}
  \end{center}
  \caption{
(a) Best electronic dielectric constant values ($\varepsilon_\text{el}$) during the optimization as a function of the number of cycles when the number of proposals in each step is different.
Here, the optimization results with \texttt{num\_search\_each\_probe = 1}, \texttt{2}, \texttt{5}, and \texttt{10} are given. TS is used, and 10 independent runs are performed; the mean and standard deviation are plotted.
Increasing the number of proposals in each step produces better $\varepsilon_\text{el}$ results with a small number of cycles.
(b) Best $\varepsilon_\text{el}$ as a function of the number of total experiments.
In terms of total experiments, the best case is using one proposal in each cycle; that is, the best case is when parallel experiments are not used.
  }
  \label{fig:bandgap_multi_probe}
\end{figure}

\subsubsection{Materials data}
The electronic dielectric constant ($\varepsilon_\text{el}$) and bandgap obtained by first-principles calculations for some semiconductors are used to demonstrate.
These data are from Ref.~\cite{PhysRevMaterials.4.103801}; the number of materials is 1,277.
Compositional descriptors generated by magpie~\cite{Ward:2016va} are used as the candidate inputs.
The descriptors generated magpie package are one of the commonly used to predict materials properties in inorganic materials. These descriptors are based on information only from the periodic table, such as atomic number, atomic weight, etc.
Although there are 132 magpie descriptors in all, some descriptors do not depend on the the data.
Thus, in this demonstration, only 108 of the descriptors are used as input.

\subsubsection{Single proposal for single-objective optimization problem} \label{sec:spsoo}

We can use a single-objective optimization problem---in this case, an optimization to obtain better materials with a higher dielectric constant---to demonstrate the efficiency of BO in PHYSBO.
Here, we consider the case in which a single input is proposed in each cycle.
For the initial data, 10 randomly sampled data values are prepared,
and 10 independent runs with different initial data are performed.
Figure~\ref{fig:bandgap_single_probe} (a) shows the average of the best $\varepsilon_\text{el}$ during the optimization as a function of the number of cycles using the various scores (random, TS, EI, and PI).
Here, `random' represents random sampling, wherein the next input is randomly selected from the input candidates in each step. As indicated, using BO enables us to identify materials with a higher $\varepsilon_\text{el}$ than when random sampling was used, even with a small number of cycles.
Additionally, in this problem, all the EI, PI, and TS methods have almost the same efficiency in terms of the number of cycles.
As already seen in Fig.~\ref{fig:scale} (b),
the TS method with the random feature map can propose the next input much faster than the EI and PI methods,
and therefore the wall-clock time to perform the BO is smaller when one uses the TS method than the EI and PI.

Next, we can consider the effect of the basis number of the random feature map on optimization performance.
Figure~\ref{fig:bandgap_single_probe} (b) shows the optimization results from BO using a random feature map $\ell$ with different basis numbers (\texttt{num\_rand\_basis} = 100, 1000, and 10000), together with the results from BO without the use of random feature map (\texttt{num\_rand\_basis} = 0).
Here, TS is used to select the next input.
As shown, when \texttt{num\_rand\_basis} is 100, the optimization performance of BO is poor relative to the other cases.
On the other hand, the performance of BO is better when \texttt{num\_rand\_basis} is 1000.

\begin{figure*}[t]
  \begin{center}
    \includegraphics[width=\textwidth]{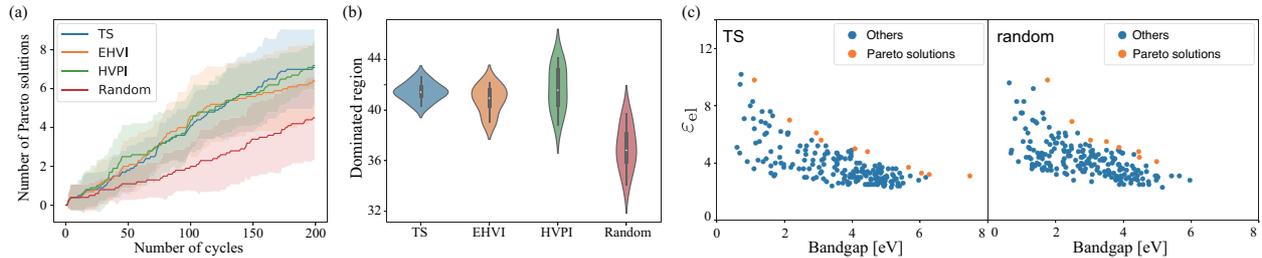}
  \end{center}
  \caption{
(a) Number of Pareto solutions found as a function of the number of cycles by PHYSBO with \texttt{score="TS"}, \texttt{"EHVI"}, and \texttt{"HVPI"} for a multi-objective optimization problem when the target properties are $E_g$ and $\varepsilon_\text{el}$.
The results are compared with random sampling.
Ten independent runs are performed; the mean and standard deviation are plotted.
(b) Dominated region depending on the method when the number of cycles is 200.
Violinplot is drawn using 10 independent runs.
(c) Examples of sampling results for TS (left figure) and `random' (right figure).
The orange points are the pareto solutions found; the blue points are other sampling points.
  }
  \label{fig:multi-objective}
\end{figure*}

\subsubsection{Multiple proposals for single-objective optimization problem}\label{sec:multi-proposals}

We can also address the effect of the number of proposals in BO.
The same single-objective optimization problem discussed in Sec.~\ref{sec:spsoo} is treated here.
Figure~\ref{fig:bandgap_multi_probe} (a) shows the average of the best $\varepsilon_\text{el}$ during the optimization as a function of the number of cycles when the number of proposals in each step is different (e.g., \texttt{num\_search\_each\_probe = 1}, \texttt{2}, \texttt{5}, and \texttt{10}).
As indicated, by increasing the number of proposals, materials with larger electronic dielectric constant values can be found more quickly.
In terms of the number of cycles, $\varepsilon_\text{el}$ is slowly growing for the single proposal case among the other BO runs.
A similar plot featuring the total number of experiments is shown in Figure~\ref{fig:bandgap_multi_probe} (b).
As can be seen here, increasing \texttt{num\_search\_each\_probe} substantially increases the number of experiments necessary to obtain a better material. 

To summarize, multiple proposals can reduce the total number of cycles (corresponding to the optimization time), 
but the total number of experiments/simulations (corresponding to the cost of optimization) is increased.
In other words, when it is effective to perform experiments/simulations in parallel, multiple proposals are useful.
On the other hand, if the cost of a single experiment/simulation is high, 
using a single proposal will identify a better input at a lower cost.

\subsubsection{Multi-objective optimization problem}

A multi-objective optimization problem in which a search for materials with both a higher bandgap $E_g$ and higher dielectric constant $\varepsilon_\text{el}$ can be used to further establish PHYSBO's efficiency.
Figure~\ref{fig:multi-objective} (a) shows the number of Pareto solutions found as a function of the number of cycles.
For BO, three types of scores for multi-objective optimization (i.e., \texttt{score="TS"}, \texttt{"EHVI"}, and \texttt{"HVPI"}) are tested, and in each cycle, a single proposal is obtained. The results show that BO was able to find many more Pareto solutions than was the case for random sampling.

To evaluate the Pareto solutions, the area of the dominated region is calculated for the results with 200 cycles.
Here, the solution space is given as a rectangle $(E_g/\text{eV}, \varepsilon_\text{el}) \in [0,14] \otimes [0, 8]$.
If the dominated region is large, many better Pareto solutions are obtained.
The violinplot for this area using 10 independent runs is shown in  Figure~\ref{fig:multi-objective} (b).
As shown, BO produces a much larger dominant area than random sampling; moreover, TS produces a better Pareto solution in the most stable way.
For reference, examples of the sampling results for TS and random sampling are shown in Figure~\ref{fig:multi-objective} (c).
Here, we have plotted the results of the case where the number of pareto solutions found (orange points) is largest for 10 independent runs. 
It can be seen that the number of Pareto solutions obtained by TS is larger than in the case of random sampling.

\subsection{Model parameter estimation}

BO can be also applied to model parameter estimation~\cite{Tamura0193785}.
As an example, we use the estimation of the magnetic interactions of the spin-$1/2$ Heisenberg chain model on 12 sites from a magnetization curve.
Three types of magnetic interactions are considered:
nearest neighbor $J_1$, next-nearest neighbor $J_2$, and 3rd-nearest neighbor $J_3$.
The Hamiltonian is written as
\begin{eqnarray}
\mathcal{H} = \sum_{i=1}^{12} 
J_1 \mathbf{S}_i \cdot \mathbf{S}_{i+1}
+J_2 \mathbf{S}_i \cdot \mathbf{S}_{i+2}
+J_3 \mathbf{S}_i \cdot \mathbf{S}_{i+3},
\end{eqnarray}
where $\mathbf{S}_i$ is the vector of the spin operator at $i$-th site and $\mathbf{S}_i = \mathbf{S}_{i+12}$.

In the model parameter estimation described here,
the target magnetization curve $\{m^{\rm target}(h_j) \}_{j=1,...,N_h}$ is given first.
$h_j$ is the magnetic field and $N_h$ is the number of magnetizations with different $h_j$.
When the interaction values $(J_1, J_2, J_3)$ are given,
the magnetization curve can be calculated by simulation as $\{m (h_j; J_1, J_2, J_3) \}_{j=1,...,N_h}$.
Thus, the difference between the target and calculated magnetization curves is given as
\begin{eqnarray}
\Delta = \sum_{j=1}^{N_h} [m^{\rm target} (h_j) - m (h_j; J_1, J_2, J_3)]^2.
\end{eqnarray}
By searching for $(J_1, J_2, J_3)$ so that $\Delta$ becomes minimum, we can estimate the model parameters that describe the target magnetization curve.
In an actual problem involving model parameter estimation~\cite{PhysRevB.101.224435},
the target property to be fitted is mainly the experimental result.
Here, however, since the aim is to investigate the efficiency of PHYSBO for model estimation, we set the target magnetization curve as calculated using the Hamiltonian with $(J_1, J_2, J_3) = (1.0, 0.5, 0.3)$; that is, these values are the solution to the model estimation problem. The magnetization curve is calculated by the $\mathcal{H}\Phi$ package~\cite{KAWAMURA2017180,web_Hphi}, which is a quantum lattice solver using the exact diagonalization method.

\begin{figure}[t]
  \begin{center}
    \includegraphics[width=0.4\textwidth]{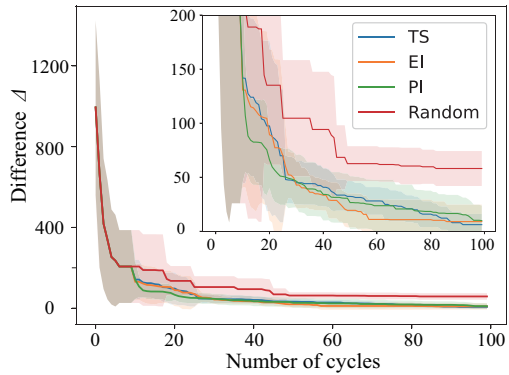}
  \end{center}
  \caption{
Best difference \tr{$\Delta$} during the optimization as a function of the number of cycles by PHYSBO using \texttt{score="TS"}, \texttt{"EI"}, and \texttt{"PI"}. The results are compared with the results of random sampling.
Ten independent runs were performed; the mean and standard deviation are plotted.
The inset is an enlarged view.
  }
  \label{model_estimation}
\end{figure}

To perform BO, the input candidates are prepared in advance.
The search space is defined as a grid in which each magnetic interaction is discretized between 0 and 2 in increments of 0.1.
The total number of input candidates is $21^3=$9,261.
Figure~\ref{model_estimation} shows the best difference during the optimization as a function of the number of cycles.
Note that this is a single-objective optimization problem and that a single proposal is used.
In addition, in PHYSBO, the input that yields the largest output is selected as the next input.
Thus, it is necessary that a negative difference is given as the output.
By using BO, we were able to find better parameters for an effective model than were produced by random sampling, even with a small number of cycles.

\section{Summary}\label{Sec:Summary}

PHYSBO is an effective Bayesian optimization package that can reduce computation time by using a random feature map, Thompson sampling, Cholesky decomposition, and parallel computing.
The PHYSBO package can treat both single- and multi-objective optimization problems.
Importantly, multiple inputs can be suggested at each cycle of the Bayesian optimization.
PHYSBO was applied to an illustrative problem involving materials screening and model estimation in order to confirm its effectiveness in the field of materials science and physics.
The results indicate that the  PHYSBO package will be a valuable tool for solving complex optimization problems in basic science.

\section*{Acknowledgments}
We wish to thank Fumiyasu Oba, Hidenori Hiramatsu, and Naoki Kawashima for their highly useful discussions. 
PHYSBO was developed under the support of the ``Project for advancement of software usability in materials science'' in fiscal year 2020 by the Institute for Solid State Physics, University of Tokyo. This work is partially supported by JSPS KAKENHI Grant No.~20H01850,  20K20522 21H01041, and 21H01008.

\bibliographystyle{elsarticle-num}
\bibliography{physbo.bib}

\end{document}